\documentclass{PoS}

\usepackage[]{latexsym}
\usepackage[]{amsfonts}
\usepackage[]{amsmath}
\usepackage[]{graphicx}
\usepackage[]{epsf}

\PoS{PoS(LAT2005)174}

\newcommand{\Tr}{\textrm{Tr}}

\newcommand{\ev}[1]{\langle #1 \rangle}
\newcommand{\myvec}[1]{{\bf #1}} 
\newcommand{\dd}{\textrm{d}}
\newcommand{\LE}{\mathcal{L}_{\textrm{E}}}
\newcommand{\SE}{S_{\textrm{E}}}
\newcommand{\MSb}{\overline{\textrm{MS}}}
\newcommand{\Nc}{N_{\rm c}}
\newcommand{\BG}{B_{\textrm{G}}}
\newcommand{\mG}{m_{\textrm{G}}}

\title{Non-perturbative plaquette in 3d pure SU(3)}

\ShortTitle{Non-perturbative plaquette in 3d pure SU(3)}

\author{\speaker{Ari Hietanen}\thanks{Supported by 
        the Magnus Ehrnrooth foundation.}\\
        Theoretical Physics Division, Department of Physical Sciences, \\ 
        P.O.Box 64, FI-00014 University of Helsinki, Finland\\
        E-mail: \email{ari.hietanen@helsinki.fi}}

\author{Keijo Kajantie\thanks{Partly supported by Academy of Finland contract no. 77744.}\\
        Theoretical Physics Division, Department of Physical Sciences, \\ 
        P.O.Box 64, FI-00014 University of Helsinki, Finland\\
        E-mail: \email{keijo.kajantie@helsinki.fi}}

\author{Mikko Laine\\
        Faculty of Physics, University of Bielefeld, 
        D-33501 Bielefeld, Germany\\
        E-mail: \email{laine@physik.uni-bielefeld.de}}

\author{Kari Rummukainen\thanks{Partly supported by Academy of Finland contract no. 104382.}\\
        Department of Physics, University of Oulu, 
        P.O.Box 3000, FI-90014 Oulu, Finland, and \\
        Department of Physics, Theory Division, CERN, 
        CH-1211 Geneva, Switzerland\\
        E-mail: \email{kari.rummukainen@cern.ch}}

\author{York Schr\"oder\\
        Faculty of Physics, University of Bielefeld, 
        D-33501 Bielefeld, Germany\\
        E-mail: \email{yorks@physik.uni-bielefeld.de}}

\abstract{We present a determination of 
  the elementary plaquette and, after the subsequent 
  ultraviolet subtractions, of the finite part of the gluon condensate,
  in lattice regularization in three-dimensional pure SU(3) gauge theory. 
  Through a
  change of regularization scheme to $\overline{\mbox{\rm MS}}$ and a matching
  back to full four-dimensional QCD, this result determines the first
  non-perturbative contribution in the weak-coupling expansion of hot QCD
  pressure. } 

\FullConference{XXIIIrd International Symposium on Lattice Field Theory\\
                 25-30 July 2005\\
                 Trinity College, Dublin, Ireland}

\begin{document}

%
\section{Introduction}

The asymptotic freedom of QCD guarantees a small coupling constant $g$ at 
large temperatures $T$. While observables can be expressed in a generalized 
power series in $g$, the loop expansion is not applicable to an 
arbitrary order in  $g$, 
because of the so-called ``infrared wall'', as pointed out by 
Linde \cite{linde} (see also ref. \cite{gpy}). 
For every observable there exists an order of the perturbative
expansion to which an infinite number of Feynman diagrams contributes. 
For the pressure this order is $g^6 T^4$. 

No way of resumming these infinitely many diagrams has been found, 
so a different approach is needed. The problem arises due to infrared
divergences in the dynamics of zero Matsubara frequency modes of gauge 
fields. Because these modes are three-dimensional (3d) we can construct 
an effective 3d pure gauge theory called Magnetostatic QCD (MQCD) which 
accounts for the non-perturbative contribution \cite{dr,generic,bn}. 
QCD and MQCD can be matched to each other by using perturbation theory.

To obtain the non-perturbative contribution we perform lattice 
measurements in MQCD~\cite{plaquette}. The observable we consider is 
the elementary 
plaquette expectation value. The theory being super-renormalisable, 
one can match the lattice regularization scheme exactly to the $\MSb$ 
scheme. This requires a perturbative 4-loop computation on the 
lattice, however, which has not been completed yet: the missing
ingredient is specified below. 
(A certain perturbative 
4-loop computation in full QCD
remains also to be carried out.)

%
\section{Relation between $\MSb$ and lattice regularization schemes}

The euclidean pure SU($\Nc$) Yang-Mills action reads
\begin{equation}
  \SE=\int \! \dd^dx \, \LE, 
  \quad\quad \LE=\frac{1}{2 g_3^2}\sum_{k,l} \Tr[F^2_{kl}],
\end{equation}
where $d=3-2\epsilon$, $g_3^2$ is the gauge coupling, $k,l=1,\dots,d$,
$F_{kl}=i[D_k,D_l]$, $D_k=\partial_k-iA_k$, $A_k=A_k^aT^a$, 
and $T^a$ are hermitean
generators of SU($\Nc$) normalised such that $\Tr[T^aT^b]=\delta^{ab}/2$.
The vacuum energy density is defined as
\begin{equation}
  f_{\MSb}\equiv-\lim_{V\rightarrow\infty}
  \frac{1}{V}\ln\left[\int\!\mathcal{D}A_k\,
  \exp\left(-\SE\right)\right]_{\MSb},
\end{equation}
where $V$ denotes the $d$-dimensional volume. The use of the $\MSb$ dimensional
regularization scheme removes any $1/\epsilon$ poles from the
expression. In fact, using dimensional regularization the 
perturbative result vanishes,
because there are no mass scales in the propagators. 
However, for dimensional reasons, the non-perturbative form of the answer is
\begin{equation}
  f_{\MSb}=-g_3^6\frac{d_A\Nc^3}{(4\pi)^4}
 \left[\left(\frac{43}{12}-\frac{157}{768}\pi^2\right)
 \ln\frac{\bar{\mu}}{2\Nc g_3^2}
  + \BG+\mathcal{O}(\epsilon)\right],
 \label{MSb}
\end{equation}
where $d_A = \Nc^2-1$.
The logarithmic term has been calculated by introducing a 
mass scale $\mG^2$ for gluon and
ghost propagators and sending $\mG^2\rightarrow 0$ after the
computation \cite{sun,gsixg}. 

Using standard Wilson discretization, 
we can write the same theory on the lattice as
\begin{equation}
  S_a=\beta \sum_{\myvec{x}}\sum_{k<l}
  \left(1-\frac{1}{\Nc}\textrm{Re}\Tr[P_{kl}(\myvec{x})]\right),
\end{equation}
where $P_{kl}$ is the plaquette, $a$ is the lattice spacing and $\beta \equiv
2\Nc/(ag_3^2)$. Hence the continuum limit is taken by $\beta \rightarrow
\infty$.  Dimensionally, the vacuum energy density consists
of terms of the form $g_3^{2n}a^{n-3}$. Thus, approaching the continuum 
limit, we can
relate $f_a$ and $f_{\MSb}$ as follows:
\begin{eqnarray}
  \Delta f & \equiv &  f_a-f_{\MSb} \label{deltaf}\\
  & = & C_1\frac{1}{a^3}\left(\ln\frac{1}{ag_3^2} +
  C_1'\right)+C_2\frac{g_3^2}{a^2}+C_3\frac{g_3^4}{a} + 
  C_4g_3^6\left(\ln\frac{1}{a\bar\mu}+C_4'\right)+\mathcal{O}(g_3^8a).
\end{eqnarray}

Taking derivatives of eq.~(\ref{deltaf}) 
with respect to $g_3^2$ and using 3d
rotational and translational symmetries on the lattice, we
obtain the master relation
\begin{equation}
  8\frac{d_A\Nc^6}{(4\pi)^4} \BG =\lim_{\beta
  \rightarrow\infty}\beta^4
  \left\{\ev{1-\frac{1}{\Nc}\Tr[P]}_a-
  \left[\frac{c_1}{\beta}+\frac{c_2}{\beta^2}
  +\frac{c_3}{\beta^3}
  +\frac{c_4}{\beta^4}(\ln\beta+c_4')\right]\right\}.
\label{master}
\end{equation}
This quantity may be called the finite part of the gluon 
condensate in lattice regularization (in certain units).

The values of the constants $c_1,\dots,c_4'$ are trivially related to those of
$C_1,\dots,C_4'$.
When $\Nc=3$, the numerical values for the $c_i$'s are
\begin{eqnarray}
  c_1 & = & \frac{d_A}{3} \approx 2.666666667,\\
  c_2 & = & 1.951315(2),\\
  c_3 & = & 6.8612(2).
\end{eqnarray}
Here $c_1$  results from a straightforward 1-loop 
computation, while $c_2$ and $c_3$ have
been calculated in refs.~\cite{hk} and \cite{pt}, respectively.
Because there is no $\bar{\mu}$-dependence in $f_a$, the value of $c_4$ is
determined by $f_{\MSb}$ in eq.~(\ref{MSb}). Consequently,
\begin{equation}
  c_4\approx2.92942132.
\end{equation}

Because the constant $c_4' = C_4'-1/3-2\ln(2\Nc)$ is still unknown, we are not
able to fully determine $\BG$. We can, however, determine the non-perturbative
input needed for it. In order to evaluate $c_4'$ a 4-loop lattice perturbation
theory calculation is required; alternatively, it can be evaluated non-diagrammatically by means of numerical stochastic perturbation theory~\cite{dir}. 

%
\section{Lattice measurements}

\begin{figure}

  \vspace*{0.5cm}

  \centerline{
  \epsfxsize=14cm
  \epsffile{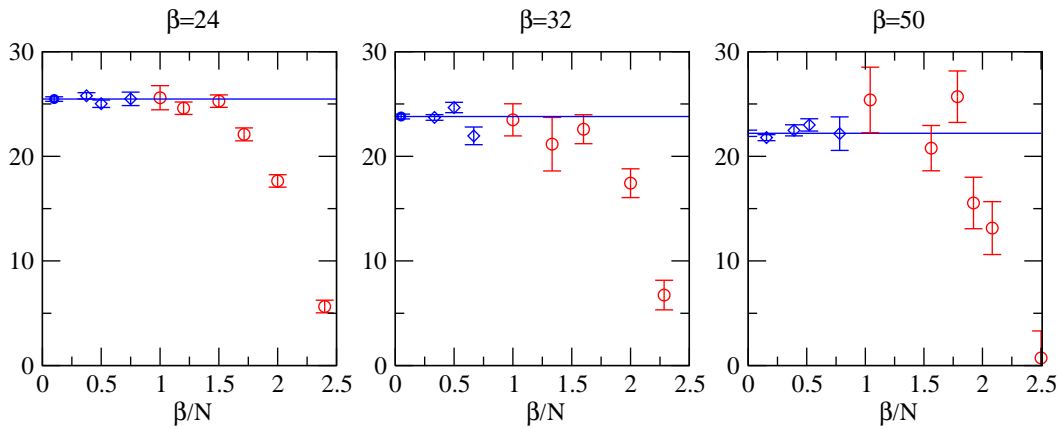}}

  \caption{Finite-volume values for
    $\beta^4\{\ev{1-\frac{1}{3}\Tr[P]}_a -
    [c_1/\beta+c_2/\beta^2+c_3/\beta^3+c_4\ln\beta/\beta^4]\}$
    as a function of the box size. The leftmost symbols indicate the
    infinite-volume estimates, obtained by fitting a constant to data in
    the range $\beta/N<1$ (blue diamonds). Points denoted by red circles
    are omitted from the extrapolation.}
  \label{ext}
\end{figure}

\begin{figure}

  \vspace*{1cm}

  \centerline{
  \epsfxsize=200pt
  \epsffile{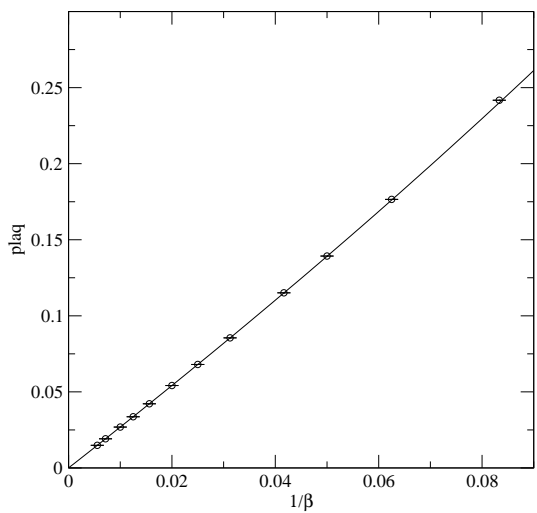}}

  \caption{The plaquette expectation value 
  plaq~$\equiv\ev{1-\frac{1}{3}\Tr[P]}_a$ as a
  function of $1/\beta$. Statistical errors are much smaller than the symbol
  sizes. The solid line contains the four known 
  terms $c_1/\beta+c_2/\beta^2+c_3/\beta^3+c_4\ln\beta/\beta^4$. The effect we
  are looking for is the difference between the data and the line.}
  \label{raw}
\end{figure}

We need the plaquette expectation value $\ev{1-\frac{1}{3}\Tr[P]}_a$ as a
function of $\beta$ so that the extrapolation $\beta \rightarrow \infty$ in
eq.~(\ref{master}) can be carried out. For each $\beta$ the infinite-volume
extrapolation is needed. Due to the non-perturbative mass gap of the theory, 
the finite-volume effects are exponentially small if the size of the box 
$L=Na$ is large compared with the inverse confinement scale $g_3^{-2}$. 
In practice finite-volume effects are invisible as soon as $\beta/N<1$. 
This is demonstrated in Fig.~\ref{ext}.
%
%
%
%
In Fig.~\ref{raw} the infinite-volume extrapolated values of
$\ev{1-\frac{1}{3}\Tr[P]}_a$ are plotted as a function of $1/\beta$.

%
%

\begin{figure}

  \vspace*{0.5cm}

  \centerline{
  \epsfxsize=200pt
  \epsffile{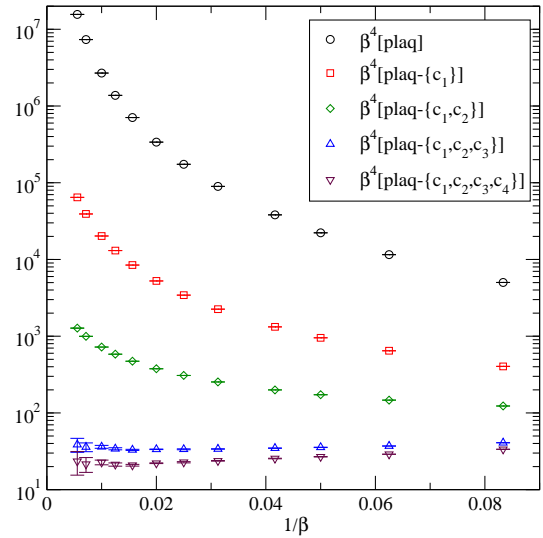}}

  \caption{The significance loss due to the subtraction of divergent lattice
  contributions. Here again plaq~$\equiv\ev{1-\frac{1}{3}\Tr[P]}_a$, and the
  symbols $c_i$ indicate which subtractions of eq.~(2.7) have been
  taken into account.}
  \label{sigloss}
\end{figure}

A major difficulty in the simulations is the significance loss caused by the
subtractions in eq.~(\ref{master}). The dominant term $c_1/\beta$ is about six
orders of magnitude larger than the effect we are interested in, namely $\sim
1/\beta^4$, if $\beta\sim 100$. Therefore the relative error of our lattice
measurements should be smaller than one part in a million.
The effect of the subtractions is illustrated in Fig.~\ref{sigloss}.

%
%

\begin{figure}

  \vspace*{0.5cm}

  \centerline{
  \epsfxsize=200pt
  \epsffile{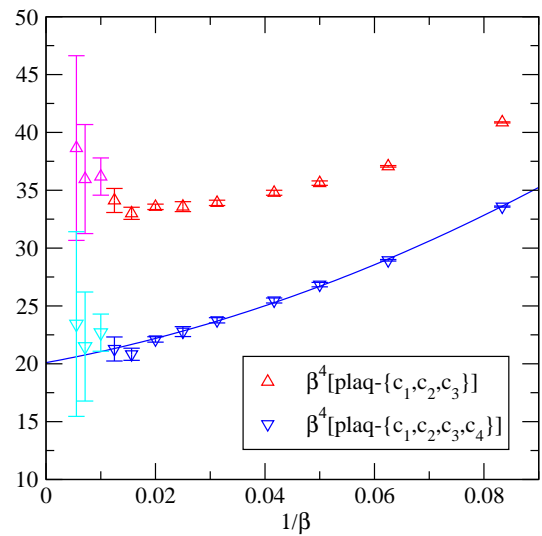}}

  \caption{The infinite-volume extrapolated data.  The effect of the 4-loop
  logarithmic divergence is to cause additional upwards ``curvature'' in the
  upper data set. The solid blue line gives the continuum 
  extrapolation. Points with
  lighter colors have so large errors that they are insignificant as far
  as the fit is concerned.}
  \label{fit}
\end{figure}

Given the infinite-volume limits, we extrapolate the data to the continuum
limit, $\beta\rightarrow\infty$. In Fig.~\ref{fit} we show two
functions:
$\beta^4\{\ev{1-\frac{1}{3}\Tr[P]}_a-
 [c_1/\beta+c_2/\beta^2+c_3/\beta^3]\}$
and
$\beta^4\{\ev{1-\frac{1}{3}\Tr[P]}_a-
 [c_1/\beta+c_2/\beta^2+c_3/\beta^3+c_4\ln\beta/\beta^4]\}$.
Even the 4-loop logarithmic term is visible in the data. For $1/\beta \le
0.01$  the significance loss grows rapidly and the error bars become quite
large, so that these data points have little effect on the fit.

%
%

The continuum extrapolation is carried out by fitting a function
$d_1+d_2/\beta+d_3/\beta^2$ to the infinite-volume extrapolated data, from 
which all the divergences ($\{c_1,c_2,c_3,c_4\}$)
have been subtracted, in the range $0.01 < 1/\beta <
0.10$. The fitted values are $d_1=20.0(7)$, $d_2=86(24)$ and $d_3=909(192)$
with $\chi^2/$dof$=5.8/6$. The error limits are the projections of the $68\%$
confidence level contour onto the various axes. The systematic errors from
the effect of higher order terms are inside these errors.

Substituting this to eq.~(\ref{master}) we obtain the final result,
\begin{equation}
  \BG+\left(\frac{43}{12}-\frac{157}{768}\pi^2\right)c_4'=
  \frac{4\pi^4}{3^6}\times20.0(7)=10.7(4).
  \label{final}
\end{equation}

%
\section{Conclusions}

We have studied the expectation value of the elementary plaquette 
in 3d pure SU(3) theory and outlined how the 3d vacuum energy density 
in the $\MSb$ scheme can be extracted from it. However, to
achieve this, the constant $c_4'$ should be determined
(cf.\ eqs.~(\ref{MSb}), (\ref{final})). This can be 
accomplished by a 4-loop matching computation, with the techniques 
discussed in ref.~\cite{dir}. 
The full QCD pressure of
order $g^6T^4$ can be obtained by a further 4-loop matching computation.

%

\end{document}